\documentclass[12pt]{iopart}

\usepackage{cite}
\usepackage{tabularx}
\usepackage{comment}
\usepackage{float}
\usepackage{braket}
\usepackage[unicode=true,bookmarks=false,breaklinks=false,pdfborder={0 0 1},colorlinks=true]{hyperref}
\hypersetup{citecolor=blue,linkcolor=blue,urlcolor=blue}
\usepackage{graphicx}
\usepackage{dcolumn}
\usepackage{bm}
\usepackage{dsfont}
\usepackage{soul}
\usepackage{xcolor}
\graphicspath{ {./plots/} }

\usepackage{iopams}
\usepackage{orcidlink}

\begin{document}
\title[Time-dependent accelerated Unruh-DeWitt detector without event horizon]{Time-dependent accelerated Unruh-DeWitt detector without event horizon}

\author{F. Sobrero$^{1}$\orcidlink{0009-0009-6497-6612}, M. S. Soares$^{1}$\orcidlink{0000-0001-5000-952X}, T.~Guerreiro$^{2}$\orcidlink{0000-0001-5055-8481} \& N. F. Svaiter$^{1}$\orcidlink{0000-0001-8830-6925}}


\address{$^{1}$ Centro Brasileiro de Pesquisas F\'isicas,  Rua Xavier Sigaud, 150 - Urca, Rio de Janeiro, RJ, Brazil }
\address{$^{2}$ Pontif\'{\i}cia Universidade Cat\'olica do Rio de Janeiro, 22451-900 Rio de Janeiro, RJ, Brazil}
\eads{\mailto{felipesobrero@cbpf.br}
, \mailto{matheus.soares@cbpf.br}, \mailto{barbosa@puc-rio.br}, \mailto{nfuxsvai@cbpf.br}}
\vspace{10pt}
\begin{indented}
\item[]January 2025
\end{indented}


\begin{abstract}
We investigate unitarily inequivalent representations of the algebra of operators in quantum field theory in the cases where there is a Fock representation of the commutation relations. We examine more closely the operational definition of a measurement device, discussing the Unruh-DeWitt and Glauber models of quantum detectors. The transition probability per unit of proper time of both detectors in different non-inertial frames of reference in Minkowski spacetime is evaluated. We first study detectors traveling in a stationary worldline with a constant proper acceleration, i.e., a hyperbolic motion, interacting with the field prepared in the Poincar\'e invariant Fock vacuum state. Next, we study the Unruh-DeWitt detector at rest in a non-uniformly accelerated frame, with a time dependent acceleration interacting with the field in the Poincar\'e invariant Fock vacuum state. We evaluate the positive frequency Wightman function for the non-uniformly accelerated frame in a finite time interval and find that it is similar to the two-point correlation function of a system in equilibrium with a thermal bath. Calculating the transition rate, the non-uniformly accelerated Unruh-DeWitt detector can be excited even if the scalar field is prepared in the vacuum state.
\end{abstract}


\maketitle


\section{Introduction}\label{intro}

The quantum revolution revolves around a fundamental question: what constitutes a measurement in the empirical world? 
In the Copenhagen interpretation of quantum mechanics, the formalism makes predictions about events computing probabilities of response of macroscopic measuring devices \cite{rosenfeld,wheeler,haagf,peres}.
Two different strategies have been used to apply mathematical concepts in describing physical reality. One is quite pragmatic, while the other emphasizes mathematical rigor.
The approach to systematizing and bringing rigor to quantum mechanics was presented by Hilbert, von Neumann and others \cite{hilbert,von neumann,von}. 
In quantum mechanics, a physical state is a positive linear functional on the algebra of self-adjoint operators, and physical observables are represented by self-adjoint linear operators acting on the rays in the Hilbert space. In the development of the whole formalism, one important result was obtained by Stone and von Neumann \cite{stone,von2}. For the system of Heisenberg commutation relations for the generalized coordinates and momenta operators $Q_{j}$ and $P_{i}$ there exists a unique irreducible representation by operators acting in a Hilbert space, unique up to a unitary transformation \cite{faddeev}. 

On the other hand, quantum mechanics was generalized to a relativistic invariant theory and also for many-particle systems where field excitations are created and destroyed. Quantum field theory, which unifies quantum mechanics with classical field theory, models dynamical systems with an uncountable number of degrees of freedom, where observables are described using a non-commutative algebra \cite{nishijima,pauli,wentzel}.
Since the Stone-von Neumann uniqueness theorem is not valid in this case, the consequences of this fact in the whole formalism of the theory of quantized fields must be discussed \cite{garding}. An example of a unitarily inequivalent representation of the canonical commutation relations was discussed by Friedrichs, searching for 
a criterion that would distinguish between field theories that allows or not a particle interpretation  \cite{myriotic1}. 
This 
representation 
is known in the literature as myriotic fields \cite{alexanian}.  These myriotic fields do not posses the no-particle state, i.e., the familiar vacuum state. Actually, the total number of excitations is infinite in 
every state of a myriotic field. This discussion led us to question the meaning of particles and measurements in relativistic theories and local quantum field theory \cite{wightschw,haagswieca,m1,m2,m3}.

In general, in quantum field theory there are infinitely many inequivalent representations of some operator algebra, with disjoint Hilbert spaces. Another example of an inequivalent representation of the operator algebra, with a distinct Hilbert space, is related to Haag’s theorem \cite{haag}. See also \cite{hove}. Although the Fock representation works very well for free fields, this representation fails for interacting fields. 
No field theory exists which differs from that of a free field \cite{greenberg1,lopuszanski,lupher}. There are two different ways to circumvented the Haag theorem. The first one is to employs a non-Fock representation in describing interacting fields, as for example the coherent states \cite{roy2}. 
Also one should consider that the time evolution of quantum fields is only locally unitarily implementable \cite{guenin}. 
Another situation with inequivalent representations of the operator algebra with distinct Hilbert spaces 
is related to 
spontaneous breakdown of continuous internal symmetries, where an arbitrary large of number of massless excitations appears \cite{jona,go,umezawa1,umezawa2,ghk,umezawa}. The excitation spectrum of a physical system is in the foundations of the formulation of quantized field theories. It was then slowly realized that different representations of the operator algebra with distinct Hilbert spaces can give quite different spectra for inequivalent representations of the operator algebra. 

Parallel to these developments, the limits of applicability of quantum field theory was put to the test also by the formulation of quantum field theory in curved space-time where problems of different nature appears \cite{TQC1, TQC2,TQC3,scd,frolov,ru}. In a curved spacetime it is possible to have incomplete geodesics with limiting value of the affine parameter, defining a geometrical singularity \cite{geroch,ryan}. Also,
in a generic pseudo-Riemannian manifold it is not possible to define positive and negative frequency modes in the Fourier expansion of field operators. Therefore in principle there are inequivalent representations of the operator algebra with distinct Hilbert spaces, each of them describing a specific physical situation.

The main purpose of the present paper is to investigate unitarily inequivalent representations of the canonical commutation relations and the response functions of different models of quanta detector in Minkowski spacetime. What comes to mind first is the existence of the Unruh-Davies effect \cite{davies11,unruh} in the situations outside of its original idea \cite{hu}.  This effect is characterized by: (i) the excitation rate is different from zero even if the apparatus is coupled to a quantum field in the vacuum state, (ii) the response of the detector is the same as it would be immersed in a bath at finite temperature, due to the presence of event horizon. Nonetheless, in a realistic scenario, a detector undergoes acceleration for a finite time. Therefore, the transition probability between the initial and final states of the detector must be evaluated over a finite time interval \cite{nami1}. We first discuss the Glauber and the Unruh-DeWitt model detectors traveling in an uniformly accelerated worldline. For large time intervals, both detector's response is similar to that a system immersed in a thermal bath at finite temperature, due to the presence of an event horizon. Next, we consider a family of detectors with non-uniform acceleration such that in the asymptotic past and future the acceleration goes to zero \cite{nu1,nu2,nu3}. We evaluate the excitation rate of the Unruh-DeWitt detector where the apparatus is coupled to a quantum field in the vacuum state. In a finite time interval, the positive frequency Wightman function is similar to the two-point correlation function of a system in equilibrium with a thermal bath. Also, the response function of non-uniformly accelerated detector and for an uniformly accelerated detector are the same for a finite time interval \cite{nami1}. Both have the same transient contributions. For large time intervals, in the case of the uniformly accelerated detector these transient terms vanish, this is exactly the manifestation of the Unruh-Davies effect. For non-uniformly accelerated detectors, we can not interpret that the detector behaves as immersed in a thermal bath, since we are not able to disregard the transient contributions. We can interpret the absence of a measured temperature as the consequence of the lack of the event horizon. 

Otherwise, is important to mention that our work shares conceptual similarities with results arising from analogue models, systems that mimic quantum field phenomena in curved spacetime. For instance, the moving mirror framework employed in \cite{mirror} captures black hole evaporation and backreation effect through classical mirror trajectories and also produces radiation spectra. In cosmological setting, \cite{cosmol} shows that spacetime expansion alone can modulate vacuum fluctuations to affect quantum communication capacities even without real particle exchange. These examples suggest that thermal or quasi-thermal responses may emerge generically from classical or quasi-static distortions of quantum fields, without the need for global horizons. In this sense, our results support the view that the excitation of an accelerated detector may be interpreted as part of a broader class of phenomena in which non-inertial motion or background geometry induces an effective thermal structure in quantum field correlations.

The structure of this work is as follows.
In section \ref{ha} some results in quantum field theory and model detectors are discussed. Section \ref{acc} presents the response function of the Glauber and Unruh-DeWitt detector models for uniformly accelerated worldlines. In section \ref{km} the response function of an Unruh-DeWitt detector at rest in a non-uniformly accelerated frame of reference is discussed.  
Conclusions are given in Sec. \ref{conc}. 
We use the units where $\hbar=c=k_{B}=1$.

\section{The postulates of quantum field theory and model detectors}\label{ha}

A generic local quantum field theory is defined in the four dimensional spacetime. The formalism is based in the following postulates: (i) state vectors associated with a physical system form a separable, normalizable Hilbert space with (in principle) a positive definite metric, (ii) covariance of the fields under Lorentz transformations and spacetime translations, i.e., there is a continuous unitary representation of the Poincar\'e group in the space of physical states of the theory, (iii) positive energy, i.e., the spectra of the generators of translations lies within the closed forward light cones, $p^{2}\geq 0$, $E\geq 0$, (iv) local commutativity or anti-commutative of Heisenberg fields or Einstein causality. This is related to the observed macroscopic causal structure in spacetime of events and finally, (v) uniqueness of the vacuum with a particle interpretation  with for example, the Fock representation \cite{fock,cook}. 
With a continuous unitary representation of the Poincar\'e group $U(a,\Lambda)$, where $a$ is a spacetime translation and $\Lambda$ is a Lorentz tranformation, the vacuum state should be invariant under $U(a,\Lambda)$ i.e., $U(a,\Lambda)|\Omega_{0}\rangle=|\Omega_{0}\rangle$. In the literature it is called the Poincar\'e invariant Fock vacuum state, naturally constructed by inertial observers.
This state is invariant under all other symmetry transformations, for instance for continuous symmetry group and discrete symmetries. Inertial observers construct the state which is a superposition of one-excitation states with all possible momenta given by $\phi(t,x)|\Omega_{0}\rangle$.  The quantum of the field interpretation is that an element of the Hilbert space represents a state of a single quantum and the Fock space is constructed representing some superposition of $n$-excitations states for various values of $n$.

The uniqueness of the Poincar\'e invariant Fock vacuum state for field theories constructed by a family of inertial observers in Minkowski spacetime is related to unitary equivalence of theories. For two theories with Hilbert spaces ${\cal{H}}_{1}$ and ${\cal{H}}_{2}$, continuous unitary representations of the Poincar\'e group in the space of physical states $U_{1}$ and $U_{2}$,  vacuum states $|\Omega_{1}\rangle$ and $|\Omega_{2}\rangle$ with fields $\varphi_{1}$ and $\varphi_{2}$ are equivalent if there is a unitary mapping $V$ of the ${\cal{H}}_{1}$ onto ${\cal{H}}_{2}$ such that $U_{1}=V^{-1}U_{2}V$, $|\Omega_{2}\rangle=V|\Omega_{1}\rangle$ and $\varphi_{1}=V^{-1}\varphi_{2}V$. 
For an hermitian scalar field of mass $m_{0}$ a representation of the canonical commutation relation is equivalent to the Fock representation if and only if the number operator exists as a densely-defined self-adjoint positive operator in the representation.  
In the canonical formalism with the Fourier expansion of free or interacting fields and the construction of the Hilbert space of states, i.e., the Fock space, the concept of a particle localized in space disappears. Any physical state of the system that belongs to a separable Hilbert space is a global object. Particle localization in field theory is realized by the detection performed by the measuring device. This raises the question. What is a detector in quantum field theory? For instance, we are assuming that the Fock vacuum state $|\Omega_{0}\rangle$ is defined by the Fourier expansion of the field operator where positive frequency modes are defined with respect to some time translation Killing field. If the worldline of the model detector is an orbit of this Killing vector, the time measured by the detector is the proper time along its worldline. For inertial worldlines the positive frequency modes in the Fourier expansion are also positive frequency modes with respect to the detector's proper time. This balance is broken for accelerated detectors. 

To define what is meant by a quanta counter, Glauber, in discussing the detection of photons, defines a detector as a system for which the rate of clicks vanishes for the Fock vacuum state \cite{glauber}.
The same definition was used by Gardiner and Zoller \cite{gz}, see also \cite {Mandel,steinmann}. The idea behind this definition is the absorptive characteristic of the detector. The intrinsic amplification effect that explains the recording of the state of the field in a classical apparatus requires taking one electron from a bound state in an atom to the continuum. A simpler model of detector is a quantum mechanical system which makes a transition from a lower to an excited state by the interaction with the field. This is the Unruh-DeWitt model detector \cite{witt,wald, JLTP}, which has a quite different conceptual framework with respect to Glauber. 
As was discussed by some authors, the Unruh-DeWitt detector is not a ideal quanta counter, it is a fluctuometer, akin to a homodyne detector \cite{homodyne}. The detector response function is the Fourier transform of the positive frequency Wightman function, evaluated in the detector worldline. Therefore this model detector measures the correlation between vacuum fluctuations along the detector worldline \cite{candelas, hinton,costa,s1}. 

In the next section we discuss the response function of both detectors  
in a non-inertial frame of reference. The
detectors with constant proper acceleration, traveling in a stationary worldline, i.e., an hyperbolic motion.

\section{The uniformly accelerated Glauber and Unruh-DeWitt detector}\label{acc}

Suppose an observer at rest in an inertial frame of reference in Minkowski spacetime. In order to canonical quantize the Hermitian scalar field we impose the operator algebra and define
$\omega_{\bold{k}}=\sqrt{\bold{k}^{2}+m_{0}^{2}}$. Next we decompose the scalar field operator into a sum of positive and negative frequency contributions  given by 
\begin{equation}
\varphi(t,\bold{x})=\varphi^{(+)}(t,\bold{x})+\varphi^{(-)}(t,\bold{x})
\end{equation}
where
\begin{equation}
\varphi^{(+)}(t,\bold{x})=\frac{1}{(2\pi)^{3/2}}
\int\frac{d^{3}k}{\sqrt{2\omega_{\bold{k}}}}a({\bold{k}})e^{-i(\omega_{\bold{k}}t-{\bold{k}}.\bold{x})}
\end{equation}
and
\begin{equation}
\varphi^{(-)}(t,\bold{x})=\frac{1}{(2\pi)^{3/2}}
\int\frac{d^{3}k}{\sqrt{2\omega_{\bold{k}}}}a^{\dagger}({\bold{k}})e^{i(\omega_{\bold{k}}t-{\bold{k}}.\bold{x})}.
\end{equation}
The annihilation and creation operators for quanta of the field $a({\bold{k}})$ and $a^{\dagger}({\bold{k}})$ satisfy the usual commutation relations. The Poincar\'e invariant Fock vacuum state $|\Omega_{0}\rangle$  is the state in which quanta are absent, i.e.  $a({\bold{k}})|\Omega_{0}\rangle=0 ~\forall ~\bold{k}$.

Consider a family of observers at rest in non-inertial reference frames with rectilinear uniformly accelerated motion. From the usual Cartesian coordinates $x^{\mu}=(t,x^{1},x^{2},x^{3})$ one defines the Rindler coordinates $y^{\mu}=(\eta,\xi,y^{2},y^{3})$ using 
\begin{equation}
t=\xi\,\sinh \eta, ~~~x^{1}=\xi\,\cosh \eta,
\end{equation}
for $x^{2}=y^{2}$ and $x^{3}=y^{3}$. Note that we have $0<\xi<\infty$ and $-\infty<\eta<\infty$. An observer travelling in the worldline $\xi=\frac{1}{a}$ has proper acceleration given by $a$. This coordinate system cover only a wedge of the Minkowski spacetime, i.e., the region $|x^{1}|>t$. Consider a congruence of wordlines, and a three-dimensional space-like hypersurface that intersects the congruence. Therefore, each observer worldline  is an orbit of a global timelike Killing vector field $\frac{\partial}{\partial\eta}$. This manifold has bifurcate Killing horizon, i.e., a pair of interconnecting null hypersurfaces that are orthogonal to the Killing field.

Let us define the vectors ${\bold{q}}=(k_{y},k_{z})$ and 
$\bold{y}=(y,z)$. The field operator can be written as a sum of positive and negative frequency contribution with respect to the Killing vector field $\frac{\partial}{\partial\eta}$ as 
\begin{equation}
\varphi(\eta,\xi,\bold{y})=
\varphi^{(+)}(\eta,\xi,\bold{y})+\varphi^{(-)}(\eta,\xi,\bold{y}).
\end{equation}
Defining $m^{2}=m_{0}^{2}+{\bold{q}}^{2}$ we have 
\begin{equation}
\varphi^{(+)}(\eta,\xi,\bold{y})
= \frac{1}{2\pi^{2}}\int_{0}^{\infty}d\nu\int d^{2}\bold{q}\sqrt{\sinh\,\pi\nu}K_{i\nu}(m\xi) b(\nu,\bold{q})\,e^{-i(\nu\,\eta-{\bold{q}}.\bold{y})},
\end{equation}
and
\begin{equation}
\varphi^{(-)}(\eta,\xi,\bold{y})
=\frac{1}{2\pi^{2}}\int_{0}^{\infty}d\nu\int d^{2}\bold{q}\sqrt{\sinh\,\pi\nu}K_{i\nu}(m\xi) b^{\dagger}(\nu,\bold{q})\,e^{i(\nu\,\eta-{\bold{q}}.\bold{y})}.
\end{equation}
where $K_{i\nu}(z)$ are the Macdonald functions of imaginary order \cite{fulling1, fulling2,raine}. The annihilation and creation operators of Rindler field quanta $b(\nu,\bold{q})$ and $b^{\dagger}(\nu,\bold{q})$ satisfy the commutation relations over a constant $\eta$
hypersurface. One can construct the Fulling Fock space, where the Fulling vacuum state is defined as $b(\nu,\bold{q})|\Omega_{F}\rangle=0$, $0\leq \nu<\infty$, $-\infty<k_{y}<\infty$ and $-\infty<k_{z}<\infty$.
Using the Bogoliubov coefficients one can write 
\begin{equation}
b(\nu,\bold{q})=\int d^{3}\bold{k}'\bigl(U(\nu,\bold{q},\bold{k}')a({\bold{k}}')+V(\nu,\bold{q},\bold{k}')a^{\dagger}({\bold{k}}')\bigr),
\end{equation} 
where 
\begin{equation}
U(\nu,\bold{q},\bold{k}')=\frac{\delta(\bold{q} - \bold{q}')}{(2\pi\omega_{\bold{k}'}(1-e^{-2\pi\nu}))^{1/2}}\Biggl(\frac{\omega_{\bold{k}'}+|\bold{k}'|}{m}\Biggr)^{i\nu}
\end{equation}
and
\begin{equation}
V(\nu,\bold{q},\bold{k})=\frac{\delta(\bold{q} + \bold{q}')}{(2\pi\omega_{\bold{k}'}(e^{2\pi\nu}-1))^{1/2}}\Biggl(\frac{\omega_{\bold{k}'}+|\bold{k}'|}{m}\Biggr)^{i\nu}
\end{equation}
in which $\bold{k}' = (k', \bold{q}')$.

Finally, we have
\begin{equation}
\langle\Omega_{0}|b^{\dagger}(\nu,\bold{q})b(\nu',\bold{q}')|\Omega_{0}\rangle=\frac{1}{e^{2\pi\nu}-1}\delta(\nu-\nu')\delta(\bold{q}-\bold{q}').
\end{equation}
Therefore the Poincar\'e invariant Fock vacuum state $|\Omega_{0}\rangle$ is a thermal state of excitations defined by the Fulling quantization. 

The same result can be obtained using the Glauber and the Unruh-DeWitt detectors.
For an operational point of view, the worldline of a uniformly accelerated detector can not be realized as trajectories of any physical object in Minkowski spacetime. In a realistic situation a detector initially moves freely, during a finite time is accelerated and after this again is inertial. Therefore, the transition probability between the initial and final states must be evaluated over a finite time interval \cite{nami1}. It was assumed that the spatial extension of the measuring device can be neglected, i.e., the detector is point-like localized and has two discrete energy levels. Also a linear coupling between the detector and the scalar field was used. 
For the case of the monopole operator, the detector-scalar field interaction Hamiltonian is
\begin{equation}
H_{int}=c_{1}\mu(\tau)\varphi(x(\tau)),
\end{equation}
where $\mu(\tau)$ is the monopole operator with two energy levels $\omega_{e}$ and $\omega_{g}
$ and $\tau$ is the proper time of the detector. 

Let us discuss the behaviour of the detector interacting with the field in  a prepared state and the detector worldline  is an orbit of a global timelike Killing vector field $\frac{\partial}{\partial\eta}$. The composite system of apparatus plus the quantum field is initially in a pure state which is the tensor product of the field state and apparatus state. The measurement process gives the probability of transition between the levels of the detector between the time interval $\tau_{f}-\tau_{i}$. 
The initial state of the field is the Poincar\'e invariant Fock vacuum state $|\Omega_{0}\rangle$. In first-order perturbation theory, defining $w_{eg}=\omega_{e}-\omega_{g}$, $\mu_{eg}=\langle e|\mu(\tau)| g\rangle$ the probability of the detector being excited, is
\begin{eqnarray}
\fl P_{eg}(\omega_{eg},\tau_{f},\tau_{i},\xi,\bold{y}) =|\mu_{eg}|^{2}\int_{\tau_{i}}^{\tau_{f}}d\tau\int_{\tau_{i}}^{\tau_{f}}d\tau'e^{-i\omega_{eg}(\tau-\tau')}\langle\Omega_{0}|\varphi(\tau,\xi,\bold{y})\varphi(\tau',\xi,\bold{y})|\Omega_{0}\rangle.
\end{eqnarray}
The response function is written as $F(\omega_{eg},\tau_{f},\tau_{i},\xi,\bold{y})$. We write
\begin{eqnarray}
\fl F(\omega_{eg},\tau_{f},\tau_{i},\xi,\bold{y}) =\int_{\tau_{i}}^{\tau_{f}}d\tau\int_{\tau_{i}}^{\tau_{f}}d\tau'e^{-i\omega_{eg}(\tau-\tau')} \langle\Omega_{0}|\varphi(\tau,\xi,\bold{y})\varphi(\tau',\xi,\bold{y})|\Omega_{0}\rangle.
\end{eqnarray}
We can split the response function into two contributions
$F(\omega_{eg},\tau_{f},\tau_{i},\xi,\bold{y})=F_{1}(\omega_{eg},\tau_{f},\tau_{i},\xi,\bold{y})+F_{2}(\omega_{eg},\tau_{f},\tau_{i},\xi,\bold{y})$ where 
\begin{eqnarray}
\fl F_{1}(\omega_{eg},\tau_{f},\tau_{i},\xi,\bold{y}) =\int_{\tau_{i}}^{\tau_{f}}d\tau\int_{\tau_{i}}^{\tau_{f}}d\tau'e^{-i\omega_{eg}(\tau-\tau')} \langle\Omega_{0}|\varphi^{(-)}(\tau,\xi,\bold{y})\varphi^{(+)}(\tau',\xi,\bold{y})|\Omega_{0}\rangle.
\end{eqnarray}
and
\begin{eqnarray}
\fl F_{2}(\omega_{eg},\tau_{f},\tau_{i},\xi,\bold{y}) 
=\int_{\tau_{i}}^{\tau_{f}}d\tau\int_{\tau_{i}}^{\tau_{f}}d\tau'e^{-i\omega_{eg}(\tau-\tau')} \langle\Omega_{0}|\varphi^{(+)}(\tau,\xi,\bold{y})\varphi^{(-)}(\tau',\xi,\bold{y})|\Omega_{0}\rangle.
\end{eqnarray}
Before continuing, let us discuss the case of an inertial detector with the response function for a finite time interval given by $F_{1}^{(i)}(\omega_{eg},\Delta t)+F_{2}^{(i)}(\omega_{eg},\Delta t)$ where
\begin{eqnarray}
\fl F_{1}^{(i)}(\omega_{eg},\Delta t) =\frac{\Delta t}{2\pi}\Biggl(-\omega_{eg}\Theta(-\omega_{eg})+\frac{\cos (\omega_{eg}\Delta t)}{\pi \Delta t}+\frac{|\omega_{eg}|}{\pi}\biggl(\mathrm{Si}(|\omega_{eg}|\Delta t)-\frac{\pi}{2}\biggr)\Biggr),
\end{eqnarray}
and
\begin{equation}
F_{2}^{(i)}(\omega_{eg},\Delta t)=
\frac{1}{2\pi^{2}}\biggl(-\gamma+\mathrm{Ci}(\omega_{eg}\Delta t)-\ln\epsilon-1\biggr),
\end{equation}
where $\gamma$ is the Euler constant and $\mathrm{Si}(z)$ and $\mathrm{Ci}(z)$ functions are the integral sine and cosine functions respectively. There are two divergences in the above equation. One given by $\ln \Delta t$ as $\Delta t\rightarrow 0^{+}$ and the other given by $\ln \epsilon$. The infinitesimal $\epsilon$ is introduced to specify correctly the singularities of the 
positive frequency Wightman function. The study of the rate of spontaneous transition eliminates the divergences of the model \cite{kullock}. Another way to avoid the divergences, where there is no switching of the interaction between the registering device and the quantum field, is to the field operator \cite{wi, jaffe,jaffe2,shilov,pct}, with test functions only in the time variable.

The asymptotic rate of spontaneous and induced emission and absorption of Rindler field quanta, i.e., the transition probability per unit proper time in the asymptotic limit $\tau_{f}-\tau_{i}\rightarrow\infty$
is given by
\begin{eqnarray}
R(\omega_{eg})=\frac{|\omega_{eg}|}{2\pi}\Biggl[&\theta(-\omega_{eg})\biggl(1+\frac{1}{e^{2\pi\sigma\omega_{eg}}-1}\biggr)+\frac{\theta(\omega_{eg})}{e^{2\pi\sigma\omega_{eg}}-1}\Biggr],
\end{eqnarray}
where $\sigma^{-1}$ is the proper acceleration of the detector.

In the Glauber theory of photodetection only the absorption term contributes to the transition rate, i.e. the transition probability per unit proper time. Also instead of considering processes involving a two energy levels it was considered  broadband detector \cite{allen}.
For the case of the Glauber model of quanta detector, the probability of the detector being excited in a generic state of the field $|\Omega_{i}\rangle$ is
\begin{eqnarray}
P_{eg}(\omega_{eg},\tau_{f},\tau_{i},\xi,\bold{y})=&& 
|m_{eg}|^{2}
\int_{\tau_{i}}^{\tau_{f}}d\tau\int_{\tau_{i}}^{\tau_{f}}d\tau'e^{-i\omega_{eg}(\tau-\tau')} \nonumber\\
&&\times  \langle\Omega_{i}|\varphi^{(-)}(\tau,\xi,\bold{y})\varphi^{(+)}(\tau',\xi,\bold{y})|\Omega_{i}\rangle.
\end{eqnarray}
where $m_{eg} = \bra{e}m\ket{g}$.

It is possible to calculate the transition probability per unit proper time of the uniformly accelerated Glauber detector, travelling in an hyperbolic worldline interacting with the scalar field prepared in the Poincar\'e invariant
Fock vacuum state $|\Omega_{0}\rangle$. This is an important result. Since the Fock vacuum state $|\Omega_{0}\rangle$ is a state with infinitely many Rindler quanta, an inequivalence among Hilbert spaces, we recover the Unruh-Davies effect using a Glauber model of quanta 
detector. The presence of the horizon with the Fulling quantization guarantees a thermal spectrum.
We get 
\begin{equation}
\frac{dP(\xi,\bold{y})}{d\tau}= \frac{1}{4\pi^{4}}\int d^{2}\bold{q}\int_{0}^{\infty}d\nu \frac{\sinh \pi\nu}{e^{2\pi\nu}-1} K_{i\nu}^{2}\biggl(\xi\sqrt{m_{0}^{2}+\bold{q}^{2}}\biggr).
\end{equation} 
We have the excitation of the Glauber detector with absorption  of a Rindler quantum from the scalar field. The above equation has a more familiar form with the choice $m_{0}^{2}=0$. In this case we get
\begin{equation}
\frac{dP(\xi,\bold{y})}{d\tau}=\frac{1}{(2\pi\xi)^{2}}\int_{0}^{\infty}d\nu\frac{\nu}{e^{2\pi\nu}-1}.
\end{equation} 
Therefore the information about the Fulling excitation spectrum of the quantum field is contained in the transition rate of the Glauber model of detector.

In the next section, we discuss the quantization performed by non-uniformly accelerated observers. The spectral density of the Unruh-DeWitt detector is presented.

\section{Unruh-DeWitt detector in non-uniformly accelerated frame}\label{km}

\begin{figure}[ht!]
    \centering
    \includegraphics[width=0.4\linewidth]{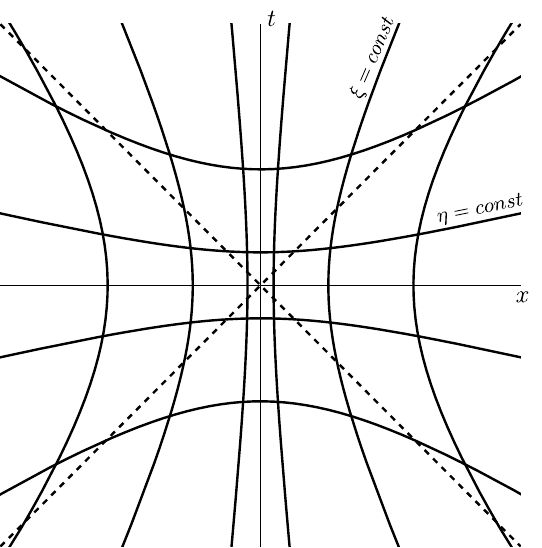}
    \caption{The curviline coordinate system defined by Eqs. \eref{eq: KM time} and \eref{eq: KM space} for constant values of $\xi$ and $\eta$. The dashed line is given by $t^2 - x^2 = 0$.}
    \label{fig: KM coordinates}
\end{figure}

The aim of this section is to evaluate the response function of the non-uniformly accelerated Unruh-DeWitt model of detector for a finite time interval. Consider a family of observers in a non-inertial reference frame, e.g., with a non-uniformly acceleration such that in the asymptotic past and future goes to zero \cite{km, miller} (see more specifically the Table 2 of Ref. \cite{miller}). Starting from the usual four-dimensional spacetime with Cartesian coordinates $x^{\mu} = (t,x^{1},y,z)$, one defines the curvilinear coordinates $x^{\mu} = (\eta,\xi,y',z')$ using
\begin{eqnarray}\label{eq: KM time}
   t &=& a^{-1} \sinh(a \eta) \cosh(a \xi), 
\end{eqnarray}  
and 
 \begin{eqnarray}\label{eq: KM space}  
   x^{1} &=& a^{-1} \sinh(a \xi) \cosh(a \eta).
\end{eqnarray}
We also have $y' = y$, $z' = z$ where $-\infty < \eta,\xi < +\infty$. The constant $\eta$ defines space-like hypersurfaces. We can see in Fig. \ref{fig: KM coordinates} that, unlike the uniformly accelerated case, this coordinate system is able to completely map Minkowski spacetime. An observer traveling in the worldline with constant $\xi$ has the proper acceleration $\alpha$ given by
\begin{equation}
    \alpha = \sqrt{2} \frac{| a \sinh(2a \xi)| }{\bigl(\cosh(2a\eta) + \cosh(2a\xi)\bigr)^{3/2}}.
\end{equation}
We immediately obtain that 
\begin{equation}
	\lim_{\eta \to \pm \infty}\alpha = 0.
\end{equation} 
The Minkowski line element in this curvilinear coordinate system is written as 
\begin{equation}
    ds^2 = A(\eta,\xi)(d\eta^2 - d\xi^2)  - dy^2 - dz^2 ,
\end{equation}
where $A(\eta,\xi)$ is written as 
\begin{equation}
	A(\eta,\xi) = \frac{1}{2}\big(\cosh(2a\eta) + \cosh(2a\xi)\big).
\end{equation}

Let us assume a massless Hermitian scalar field $\phi(\eta,\xi,y,z)$, described by the following Klein-Gordon equation:
\begin{equation}
    \Big( A^{-1}(\eta,\xi)\left( \partial^{2}_{\eta} - \partial^{2}_{\xi} \right) - \partial^{2}_y - \partial^{2}_z \Big)\phi(\eta,\xi,y,z) = 0.
\end{equation}
Using Fourier transform we define:
\begin{equation}
	\phi(\eta,\xi,y,z) = \int \frac{d^{2}\bold{k}}{2\pi}e^{-i \bold{k}.\bold{x}}\phi_{\bold{k}}(\eta,\xi),
\end{equation}
where $\bold{k}.\bold{x} = k_y y + k_z z$. Then, the above equation becomes:
\begin{equation}
\partial^{2}_\eta\phi_{\bold{k}}(\eta,\xi) - \partial^{2}_{\xi}\phi_{\bold{k}}(\eta,\xi) + \frac{1}{2}\bold{k}^2\big( \cosh(2a \eta) + \cosh(2a \xi)\big)\phi_{\bold{k}}(\eta,\xi) = 0.
\end{equation}
Next, using separation of variables, i.e., $\phi_{\bold{k}}(\eta,\xi) = \phi^{(1)}_{\bold{k}}(\eta)\phi^{(2)}_{\bold{k}}(\xi)$ we find two differential equations:
\begin{eqnarray}
	\partial^{2}_\eta\phi^{(1)}_{\bold{k}}(\eta) + \frac{1}{2}\big(\bold{k}^2 \cosh(2a \eta) - 2\lambda\big)\phi^{(1)}_{\bold{k}}(\eta) &=& 0
\end{eqnarray}
and
\begin{eqnarray}
	\partial^{2}_{\xi}\phi^{(2)}_{\bold{k}}(\xi) - \frac{1}{2}\big(\bold{k}^2 \cosh(2a \xi) + 2\lambda\big)\phi^{(2)}_{\bold{k}}(\xi) &=& 0.
\end{eqnarray}
where $\lambda$ is the separation constant. The solution of both equations are the modified Mathieu functions \cite{ab}.

In the usual quantization procedure we have that observers are moving along integral curves of a Killing vector field. The Fock space defines quantum states of excitations seen by observers in the Killing trajectories. In a generic curved space-time, in the static case we make use of the hypersurface-orthogonality of the static Killing vector to define the whole Fock space.
In the case where there is a global time-like Killing vector, but it is not orthogonal to spacelike hypersurfaces, we say the spacetime is stationary.
In stationary metrics there is an ambiguity in constructing the Fock space of the theory. There is no natural prescription for decomposing the field operator into its positive and negative frequency parts. The case of non-stationary metric is more involved.
A method of Hamiltonian diagonalization was proposed in Refs. \cite{ru1,ru2}. This method leads us to the situation where excitations are spontaneously created and annihilated, as for example in expanding universes \cite{creation1,creation2,creation3,ford}. The definition of creation and annihilation operators in this scenario is problematic since we are not able to identify positive and negative frequency modes in the set of solutions given by the modified Mathieu functions. Therefore, let us discuss only the behaviour of the Unruh-DeWitt detector at rest in this non-uniformly accelerated frame in finite times using the two-point correlation function.

We can consider the field correlation function $\langle \Omega_0|\phi\big(x(\tau')\big) \phi\big(x(\tau'')\big) |\Omega_0 \rangle$ and then calculate the response function for a non-uniformly accelerated observer. The two-point positive frequency Wightman function along the worldline of an inertial detector is:
\begin{equation}\label{eq: correlation function}
    \langle \Omega_0| \phi\big(x(t')\big) \phi\big(x(t'')\big)| \Omega_0 \rangle  = -\frac{1}{(4\pi)^2} \frac{1}{\Delta t^2-(\bold{x}'-\bold{x}'')^2 },
\end{equation}
where $\Delta t = t' - t''$.
We turn now to an observer undergoing a non-uniform acceleration with the field prepared in the Fock vacuum $|\Omega_{0}\rangle$. Therefore, considering $\xi = \xi_0$ where $\xi_0$ is a positive constant and $y = z = 0$, the coordinate $\eta = \eta(\tau)$ satisfies the equation
\begin{equation}
    \frac{1}{2}\Big( \cosh\big(2a\eta(\tau)\big) + \cosh(2a\xi_0) \Big)\dot{\eta}^{2}(\tau) = 1
\end{equation}
which can not be solved analytically, where  $\dot{\eta}=\frac{d\eta}{d\tau}$.
Considering the case where $|\eta(\tau)| \gg \xi_0$ we have that the above equation can be simplified,
\begin{equation}
    \frac{1}{2}\cosh\big(2a\eta(\tau)\big)\dot{\eta}^{2}(\tau) \approx 1, \,\,\,~|\eta(\tau)| \gg \xi_0.
\end{equation}
Even with this simplification, we can not obtain an analytical expression for $\eta(\tau)$.
Considering $a > 0$ and $|\eta(\tau)| \gg 1/a$, we get $\cosh(2a\eta(\tau)) \approx e^{\pm 2a \eta(\tau)}/2$ for $|\eta(\tau)| \gg 1/a$. Then,
\begin{equation}
    \frac{1}{4}e^{\pm2a\eta(\tau)}\dot{\eta}^{2}(\tau) \approx 1, ~|\eta(\tau)| \gg \frac{1}{a} \gg \xi_0
\end{equation}
whose solution is
\begin{equation}
    \eta(\tau) \approx \pm \frac{1}{a}\log(2a \tau). 
\end{equation}
for $\tau \gg \frac{e^{a \xi_0}}{a}$.
We obtain therefore that $t(\tau)$ and  $x(\tau)$ are written as 
\begin{eqnarray}
   t(\tau) &\approx& \pm \frac{1}{a} \left( \tau - \frac{1}{4a^2 \tau} \right) \cosh(a \xi_0)
   \label{eq: proper t-KM}
\end{eqnarray}
and   
 \begin{eqnarray}  
   x^{1}(\tau) &\approx& ~~\frac{1}{a} \left( \tau + \frac{1}{4a^2 \tau} \right) \sinh(a \xi_0). \label{eq: proper x-KM}
\end{eqnarray}
Finally replacing the above equations in Eq. \eref{eq: correlation function}, we have for  $\tau' \gg \frac{e^{a \xi_0}}{a}$ and $~ \tau'' \gg \frac{e^{a \xi_0}}{a}$ that the two-point correlation function has the form
\begin{equation}\label{eq: large times Wightman function}
    \langle \Omega_0| \phi\big(x(t')\big) \phi\big(x(t'')\big) |\Omega_0 \rangle \approx -\frac{1}{(4\pi)^2} \frac{1}{(\tau' - \tau'')^2}.
\end{equation}
This result is expected since for asymptotic limits the acceleration goes to zero and therefore the correlation function coincides with the two-point function of an observer at rest in an inertial frame of reference.

To proceed we are interested in calculating the response function of Unruh-DeWitt detector for a finite time interval during a period of non-constant acceleration when $\xi_0 \gg |\eta(\tau)|$. Using that $2a\,\xi_0 = \cosh^{-1}(2\zeta^2_0)$ (such that $a$ and $\xi_0 > 0$, then $\zeta_0 > 0$) we get 
\begin{equation}
    1 - \zeta^2_0\dot{\eta}^{2}(\tau) \approx 0,
\end{equation}
whose solution is given by
\begin{equation}
    \eta(\tau) \approx \pm\frac{\tau}{\zeta_0}, ~\tau \ll \xi_0 \zeta_0.
\end{equation}
We obtain that 
 $t(\tau)$ and $x(\tau)$ are written as
\begin{eqnarray}
   t(\tau) &\approx& \pm \frac{1}{a} \sinh\left(a\frac{\tau}{\zeta_0}\right) \cosh(a \xi_0) \label{eq: proper t-KM 2}
\end{eqnarray}
and
 \begin{eqnarray}   
   x^{1}(\tau) &\approx& ~~\frac{1}{a} \cosh\left(a\frac{\tau}{\zeta_0}\right) \sinh(a \xi_0). \label{eq: proper x-KM 2}
\end{eqnarray}
for $\tau \ll \xi_0 \zeta_0$. This approximation is leading us to a quasi-static scenario, in the sense that the acceleration varies very slowly in this time interval.
Finally replacing these previous equations in Eq. \eref{eq: correlation function} we have for
$\tau' \ll \xi_0 \zeta_0$ and $\tau'' \ll \xi_0 \zeta_0$ that the positive frequency Wightman function
is written as
\begin{eqnarray}
    \fl \langle \Omega_0| \phi\big(x(t')\big) \phi\big(x(t'')\big) |\Omega_0 \rangle   \approx -\frac{1}{(4\pi)^2}\left(\frac{a}{2\zeta_0}\right)^{2}\mathrm{csch}^2\left(\frac{a}{2\zeta_0}(\tau'' - \tau'-i\epsilon)\right),
\end{eqnarray}
the small imaginary part $i\epsilon$ was introduced to perform a contour integral of the response function. This is exactly the same correlation function for the system interacting with a thermal bath \cite{milonni2013quantum} with temperature 
\begin{equation}\label{eq: beta}
    \beta^{-1} = \frac{a}{2\pi\zeta_0},
\end{equation}
which is also the same obtained for an uniformly accelerated observer, which is possible due the fact that the acceleration is quasi-uniform in this time interval. We call this quantity, defined in Eq. \eref{eq: beta}, a local temperature because of the similarity between both Wightman functions. However, this do not means that the detector is in fact measuring a thermal bath as the case of an uniformly accelerated detector. Remembering that $2a\, \xi_0 = \cosh^{-1}(2\zeta^2_0)$, i.e., the local temperature can also be written as
\begin{equation}
    \beta^{-1} = \frac{a}{\bigl(2\pi^2 \cosh(2a \xi_0)\bigr)^{1/2}}.
\end{equation}
The response function in a finite time interval is given by
\begin{eqnarray}
\fl F(\omega_{eg},\beta,\tau_{f},\tau_{i},\xi,\textbf{0}) = -\frac{1}{(4\beta)^2}\int_{\tau_{i}}^{\tau_{f}}d\tau\int_{\tau_{i}}^{\tau_{f}}d\tau'e^{-i\omega_{eg}(\tau'-\tau)} \mathrm{csch}^2\left(\frac{\pi}{\beta}(\tau' - \tau-i\epsilon)\right),
\end{eqnarray}
where $\beta$ is defined in Eq. \eref{eq: beta}. From here, the calculation is straightforward. For more details of the procedure, see Ref. \cite{nami1}. We obtain, defining $\Delta \tau = \tau_f - \tau_i$, that $F(\omega_{eg},\beta,\tau_{f},\tau_{i},\xi,\textbf{0})$ can be written as
\begin{eqnarray}
F(\omega_{eg}, \beta, \tau_{f}, \tau_{i}, \xi, \mathbf{0}) &&= \frac{\Delta \tau}{2\pi^2} H_{(1)}(\omega_{eg}, \beta, \Delta \tau) + \frac{1}{2\pi^2}H_{(2)}(\omega_{eg}, \beta, \Delta \tau) \nonumber \\
&& + \frac{\Delta \tau}{2\pi} \left( \theta(-\omega_{eg}) \bigg( 1 - \frac{1}{e^{|\omega_{eg}|\beta} - 1} \bigg)  
+ \frac{\theta(\omega_{eg}) }{e^{|\omega_{eg}|\beta} - 1} \right) \label{eq: KM and Rindler rate}
\end{eqnarray}
where $\theta(z)$ is the Heaviside step function and
\begin{eqnarray}
H_{(1)}(\omega_{eg},\beta, \Delta \tau) 
&=& |\omega_{eg}|\bigg( \mathrm{Si}(|\omega_{eg}| \Delta \tau) - \frac{\pi}{2} \bigg) \nonumber \\
&&+ \int_{\Delta \tau}^{\infty} d\chi \cos(\omega_{eg} \chi) \left(\frac{\pi^2}{\beta^2 \sinh^2(\pi \chi/\beta)} - \frac{1}{\chi^2} \right)
\end{eqnarray}
and
\begin{eqnarray}
H_{(2)}(\omega_{eg},\beta, \Delta \tau) 
&=&  2\pi^2 \cos(\omega_{eg} \Delta \tau) + \int^{\Delta \tau}_{0} d\chi \frac{\cos(\omega_{eg} \chi) - 1}{\chi} \nonumber \\
&& + \frac{1}{\beta^2}\int^{\Delta \tau}_{0} d\chi \chi \cos(\omega_{eg} \chi) \left( \frac{\pi^2}{\sinh^2(\pi \chi/\beta)} - \frac{\beta^2}{\chi^2} \right).
\end{eqnarray}

We can define a finite rate given by
\begin{equation}
	R(\omega_{eg},\beta,\Delta \tau) = \frac{d}{d\Delta \tau} F(\omega_{eg},\beta,\tau_{f},\tau_{i},\xi,\textbf{0}).
\end{equation}
We would like to point out that there is a contribution in the result that gives the thermal properties, which can be clearly seen in Eq. \eref{eq: KM and Rindler rate}, and other terms related to the finite time evaluation given by $H_{1,2}(\omega_{eg},\beta, \Delta \tau)$. These last terms vanish when we do the limit $\Delta \tau \to \infty$, and it can be done without problem for a uniformly accelerated observer, resulting in the Unruh-Davies effect. However, the same can not be done for non-uniformly observers. This is because the calculated positive frequency Wightman function is only valid for $\tau \ll \xi_0 \zeta_0$, for large proper times, we have Eq. \eref{eq: large times Wightman function}. For this reason, even though the excitation rate of the non-uniformly accelerated detector is non-zero: the Unruh-Davies effect can not be obtained from our scenario.

\section{Conclusions}\label{conc}

In this paper we first discuss the violation of the Stone-von Neumann theorem for  systems described by uncountable number of degrees of freedom, as in quantum field theory. We discuss the deeper consequences of this fact in the whole formalism of the theory of quantized fields. For a hermitian massive scalar field, a representation of the canonical commutation relation is unitarily equivalent to the Fock representation if and only if the number operator exists as a densely-defined self-adjoint positive operator in the representation. The criterion which a quantum field theory must satisfy to allow a particle interpretation is also analised.

How does particles appear in the theory of quantum fields?  In particle physics scenario we observe particles in scattering experiments. The central operator in scattering theory is the $S$-matrix, from which transition probabilities can be computed \cite{heisenberg}. The  main idea of the $S$-matrix theory is to eliminating the concept of field in the description of elementary particles. In the remote past and remote future we are able to define free particles in these asymptotic regions. With the Yang-Feldman equations \cite{yang} we define asymptotic fields. Using the fact that asymptotic fields are free fields, one can construct the Fock representation based in the vacuum state $|\Omega_{0}\rangle$, using asymptotic completeness. Next, we define an unitary transformation $S$, called the $S$-matrix defined by $\varphi^{out}(x)=S^{-1}\varphi^{in}(x)S$. With the $S$-operator transition probabilities of particle reactions can be calculated \cite{stapp2}. The central problem of this approach is the use of analytic functions of several variables, a challenging subject since the generalization of many results of the theory of complex functions of one variable is a difficult task. 

At this stage, a remark is appropriate concerning the non-localizability of particles in quantum field theory. The mathematical description of a particle in quantum field theory is not based in localization in a space-like hypersurface, but on spectral properties, i.e., a momentum space property. In the canonical formalism with the Fourier expansion of free or interacting fields and the construction of the Hilbert space of states, i.e., the Fock space, the concept of a particle localized in space disappears. Any physical state of the system that belong to a separable Hilbert space is a global object. Particle localization in field theory is realized by the detection performed by the measuring device. This raises the question. What is a detector in quantum field theory? 

To define what is meant by a particle detection, Glauber, discussing the detection of photons, defined a quanta counter as a system for which the rate of clicks vanishes in the vacuum state. The absorptive characteristic of the detector is fundamental.
This definition is different from the Unruh-DeWitt model detector. The detector response function is defined as the Fourier transform of the two-point vacuum expectation value of the product of the field operators. As it was discussed, the Unruh-DeWitt detector is measuring the correlation between vacuum fluctuations in its proper worldline. 

For the case of detectors with constant proper acceleration, for an operational point of view, the worldline of that accelerated detector can not be realized as trajectories of any physical object in Minkowski spacetime. In a realistic situation, a detector initially moves freely, during a finite time is accelerated and after this again is inertial. The response function of the Unruh-DeWitt detector at rest in a scenario with a non-uniformly accelerated frame whose observer starts and ends inertial, is presented.  We have obtained an excitation rate different from zero even if the apparatus is coupled to a quantum field in the vacuum state with the absence of an event horizon. Although the positive frequency Wightman function in the non-uniformly accelerated frame matches that of a system interacting with a thermal bath, the presence of transient terms prevents us from interpreting the detector as being immersed in a thermal bath. This result feeds into the earlier discussion of how particles can be interpreted. Showing that even in a flat space and without an event horizon, we can construct a detector that clicks with the field prepared in the vacuum state. 

In another aspect of this discussion, it is also important to highlight the technical difficulty in measuring the Unruh effect. Considering an uniform acceleration of the order of $10^{20} m/s^2$ would take us to a temperature of the order of $1K$. In \cite{berry}, the authors show that the use of the Berry phase for this detection is quite promising and shows that accelerations $10^9$ times smaller could be detected, which is considerably better but still very problematic. One way to achieve constant and high accelerations using particles would be by introducing a charged particle into an electric field. However, to achieve the accelerations that would make it feasible to measure the Unruh effect, we would also be approaching the electric field in the regime that generates the Schwinger effect, and these two effects could be confused, as shown by Volovik in \cite{volovik}. In addition to the issues discussed previously, we still have the problem that measurements will need to be made in a finite time and this generates the transient terms \cite{nami1}, making the measurement even more difficult. Perhaps, a more feasible way to realize it is to give up a ``purely thermal'' measurement, i.e., detect the excitation of the detector without a thermal spectrum and go to a scenario that is easier to generate high accelerations without a strong electric field. A possibility is circular motion. The recent work of Unruh \cite{JLTP} gives an interesting proposal combining circular motion and interferometry that could be a good way to attack the problem and, one day, measure an excitation for a detector in the vacuum state.

We would like to point out that there is another well known example in the literature where the distortion of vacuum fluctuations gives a macroscopic measurable consequence: the Casimir effect. Quantum field boundary conditions distort the vacuum modes \cite{ca,an,plu,mi,ss1,ss2,ss3}, yielding a force that has been measured in different geometric configurations. One can expect similar effects in critical systems with infinite correlation lengths in the presence of boundaries. This is known in the literature as the critical Casimir effect \cite{cc1,dan,cc2}.

A natural continuation of this work is to study the transition rate of a Unruh-DeWitt detector using the coordinate system defined by Eq. \eref{eq: KM time} and \eref{eq: KM space} for large time intervals, leaving the Rindler scenario given by Eq. \eref{eq: proper t-KM} and \eref{eq: proper x-KM}. This topic is under investigation by the authors.

\section*{Acknowledgments}
 This work was partially supported by Conselho Nacional de Desenvolvimento Cient\'{\i}fico e Tecnol\'{o}gico (CNPq), No. 303436/2015-8, Coordenação de Aperfeiçoamento de Pessoal de Nível Superior - Brasil (CAPES) - Finance Code 001, Conselho Nacional de Desenvolvimento Científico e Tecnológico (CNPq), Fundação de Amparo \`a Pesquisa do Estado do Rio de Janeiro (FAPERJ Scholarship No. E-26/200.252/2023 and E-26/202.762/2024) and Fundação de Amparo \`a Pesquisa do Estado de São Paulo (FAPESP process No. 2021/06736-5). This work was supported by the Serrapilheira Institute (grant No. Serra - 2211-42299) and StoneLab.


\end{document}